\documentclass[10pt]{article}
\usepackage{graphicx}
\usepackage{amsmath}
\usepackage{amssymb}
\usepackage{caption2}
\begin{document}
\begin{center}
{\large {\bf \sc{  Analysis of the scalar  doubly charmed  hexaquark state  with QCD sum rules
  }}} \\[2mm]
Zhi-Gang  Wang \footnote{E-mail: zgwang@aliyun.com.  }   \\
 Department of Physics, North China Electric Power University, Baoding 071003, P. R. China
\end{center}

\begin{abstract}
In this article, we study the scalar-diquark-scalar-diquark-scalar-diquark type  hexaquark state  with the QCD sum rules by carrying out the operator product expansion up to the vacuum condensates of dimension 16.   We obtain the lowest hexaquark mass $6.60^{+0.12}_{-0.09}\,\rm{GeV}$, which can be confronted to the experimental data in the future.
\end{abstract}



\section{Introduction}

In the past years,   a number of new charmonium-like states have been observed, some are excellent candidates for the exotic states, such as tetraquark states and molecular states,     and    the
  spectroscopy of the charmonium-like states have attracted  much attentions \cite{PDG}. The QCD sum rules play  an important role  in assigning those new charmonium-like states \cite{TetraquarkQCDSR,WangHuang-3900,Wang-4660-2014}.

The scattering amplitude for one-gluon exchange  is proportional to
\begin{eqnarray}
t^a_{ij}t^a_{kl}&=&-\frac{1}{3}\left(\delta_{ij}\delta_{kl}-\delta_{il}\delta_{kj}\right)
 +\frac{1}{6}\left(\delta_{ij}\delta_{kl}+\delta_{il}\delta_{kj}\right)\, ,
\end{eqnarray}
where the $t^a$ is the generator of the $SU_c(3)$ gauge group.   The negative sign in front of the antisymmetric  antitriplet indicates the interaction
is attractive  while the positive sign in front of the symmetric sextet indicates
 the interaction  is repulsive. The
attractive interaction of one-gluon exchange  favors  formation of
the diquarks in  color antitriplet $\overline{3}_{ c}$, flavor
antitriplet $\overline{3}_{ f}$ and spin singlet $1_s$ or flavor
sextet  $6_{ f}$ and spin triplet $3_s$ \cite{One-gluon}.
The color antitriplet  diquarks $\varepsilon^{ijk} q^{T}_j C\Gamma q^{\prime}_k$
  have  five  structures  in Dirac spinor space, where $C\Gamma=C\gamma_5$, $C$, $C\gamma_\mu \gamma_5$,  $C\gamma_\mu $ and $C\sigma_{\mu\nu}$ for the scalar, pseudoscalar, vector, axialvector  and  tensor diquarks, respectively.
The calculations based on the QCD sum rules  indicate that  the favored configurations are the $C\gamma_5$ and $C\gamma_\mu$ diquark states \cite{WangDiquark,WangLDiquark}, while the heavy-light $C\gamma_5$ and $C\gamma_\mu$ diquark states have almost  degenerate masses  \cite{WangDiquark}.
We can construct the lowest tetraquark states by the  $C\gamma_5$ and $C\gamma_\mu$ diquark states and antidiquark states, for example, the $Z_c(3900)$
can be tentatively assigned to be the ground state  $C\gamma_5\otimes \gamma_\mu C-C\gamma_\mu\otimes\gamma_5 C$ type tetraquark state \cite{WangHuang-3900}.
The diquark-antidiquark type tetraquark states have been studied extensively with the QCD sum rules.

In the QCD sum rules for the four-quark states, the largest power of the QCD spectral densities $\rho(s)\propto s^4$, the integral $\int_{4m_c^2}^\infty ds \rho(s) \exp\left( -\frac{s}{T^2}\right)$ converges slowly,  the pole dominance condition is difficult to satisfy, where the $T^2$ is the Borel parameter.
In previous work, we  study the energy scale dependence of the QCD sum rules for the hidden-charm and hidden-bottom tetraquark states and molecular states
 for the first time, and suggest a  formula,
\begin{eqnarray}
\mu&=&\sqrt{M^2_{X/Y/Z}-(2{\mathbb{M}}_Q)^2} \, ,
 \end{eqnarray}
 with the effective mass ${\mathbb{M}}_Q$  to determine the energy scales of the  QCD spectral densities \cite{WangHuang-3900,Wang-4660-2014,WangHuang-NPA-2014,WangHuang-mole}, where the $X$, $Y$, $Z$ denote the tetraquark states and molecular states. The formula enhances the pole contributions remarkably.

In this article, we extend our previous work to study the scalar hexaquark state $uuddcc$ with the QCD sum rules in details. We construct the scalar-diquark-scalar-diquark-scalar-diquark type current, which is supposed to couple potentially  to the lowest hexaquark state.  In the QCD sum rules for the six-quark states, the largest power of the QCD spectral densities $\rho(s)\propto s^7$, the pole dominance condition is more difficult to satisfy compared to the QCD sum rules for the four-quark states. We use the energy scale formula to enhance the pole contributions.

The article is arranged as follows:  we derive the QCD sum rules for the mass and pole residue of  the
scalar  doubly charmed hexaquark state in Sect.2;  in Sect.3, we present the numerical results and discussions; and Sect.4 is reserved for our
conclusion.

\section{The QCD sum rules for  the  scalar  doubly charmed  hexaquark state }
In the following, we write down  the two-point correlation function $\Pi(p)$  in the QCD sum rules,
\begin{eqnarray}
\Pi(p)&=&i\int d^4x e^{ip \cdot x} \langle0|T\left\{J(x) J^{\dagger}(0)\right\}|0\rangle \, ,
\end{eqnarray}
where
\begin{eqnarray}
J(x)&=&\varepsilon^{abc}\varepsilon^{aij}\varepsilon^{bkl}\varepsilon^{cmn}\, u^{T}_i(x)C\gamma_5d_j(x) \,u^{T}_k(x)C\gamma_5 c_l(x)\, d^{T}_m(x)C\gamma_5 c_n(x) \, ,
\end{eqnarray}
 the $a$, $b$, $c$, $i$, $j$, $k$, $l$, $m$, $n$ are color indexes, the $C$ is the charge conjugation matrix. We construct   the scalar-diquark-scalar-diquark-scalar-diquark type current   $J(x)$ to interpolate the lowest
   hexaquark state $Z_{cc}^{++}$.

At the phenomenological side,  we insert  a complete set of intermediate hadronic states with
the same quantum numbers as the current operator $J(x)$ into the
correlation function $\Pi(p)$  to obtain the hadronic representation
\cite{SVZ79,Reinders85}, and isolate the ground state
contribution,
\begin{eqnarray}
\Pi(p)&=&\frac{\lambda_{Z}^2}{M^2_{Z}-p^2}  +\cdots  \, ,
\end{eqnarray}
where the pole residue  $\lambda_{Z}$ is defined by $ \langle 0|J(0)|Z_{cc}^{++}(p)\rangle=\lambda_{Z}$.

 In the following,  we briefly outline  the operator product expansion for the correlation function $\Pi(p)$ in perturbative QCD.  We contract the $u$, $d$  and $c$ quark fields in the correlation function $\Pi(p)$ with Wick theorem, and obtain the result:
 \begin{eqnarray}
 \Pi(p)&=&-i\varepsilon^{abc}\varepsilon^{aij}\varepsilon^{bkl}\varepsilon^{cmn}\varepsilon^{a^\prime b^\prime c^\prime }\varepsilon^{a^\prime i^\prime j^\prime }\varepsilon^{b^\prime k^\prime l^\prime }\varepsilon^{c^\prime m^\prime n^\prime }\int d^4x e^{ip \cdot x}   \nonumber\\
&&\left\{{\rm Tr}\left[ \gamma_{5}D_{jj^{\prime}}(x)\gamma_{5} CU^T_{ii^{\prime}}(x)C\right] {\rm Tr}\left[ \gamma_{5}C_{ll^{\prime}}(x)\gamma_{5} CU^T_{kk^{\prime}}(x)C\right]      {\rm Tr}\left[ \gamma_{5}C_{nn^{\prime}}(x)\gamma_{5} CD^T_{mm^{\prime}}(x)C\right] \right. \nonumber\\
&&-{\rm Tr}\left[ \gamma_{5}C_{ll^{\prime}}(x)\gamma_{5} CU^T_{ik^{\prime}}(x)C \gamma_5 D_{jj^\prime}(x)\gamma_5C U^T_{ki^\prime}(x)C\right]       {\rm Tr}\left[ \gamma_{5}C_{nn^{\prime}}(x)\gamma_{5} CD^T_{mm^{\prime}}(x)C\right]  \nonumber\\
&&-{\rm Tr}\left[ \gamma_{5}C_{ll^{\prime}}(x)\gamma_{5} CU^T_{kk^{\prime}}(x)C\right]
 {\rm Tr}\left[ \gamma_{5}C D^T_{mj^{\prime}}(x)C \gamma_{5}  C_{nn^{\prime}}(x) \gamma_5 C D^T_{jm^\prime}(x)C\gamma_5U_{ii^\prime}(x)\right]  \nonumber\\
 &&-{\rm Tr}\left[ \gamma_{5}D_{jj^{\prime}}(x)\gamma_{5} CU^T_{ii^{\prime}}(x)C\right]
 {\rm Tr}\left[ \gamma_{5} C_{ln^{\prime}}(x) \gamma_{5} C D^T_{mm^{\prime}}(x)C \gamma_5  C_{nl^\prime}(x) \gamma_5CU^T_{kk^\prime}(x)C \right]  \nonumber\\
 &&+{\rm Tr}\left[ \gamma_{5}C C^T_{ll^{\prime}}(x)C\gamma_{5} U_{ki^{\prime}}(x)  \gamma_{5}C D^T_{mj^{\prime}}(x)C \gamma_{5}  C_{nn^{\prime}}(x) \gamma_5 C D^T_{jm^\prime}(x)C\gamma_5U_{ik^\prime}(x)\right]  \nonumber\\
 &&+{\rm Tr}\left[ \gamma_{5} C_{ln^{\prime}}(x)\gamma_{5} CD^T_{mm^{\prime}}(x)C  \gamma_{5} C_{nl^{\prime}}(x) \gamma_{5} C U^T_{ik^{\prime}}(x)C \gamma_5  D_{jj^\prime}(x)\gamma_5CU^T_{ki^\prime}(x)C\right]  \nonumber\\
 &&+{\rm Tr}\left[ \gamma_{5} C_{ln^{\prime}}(x)\gamma_{5} CD^T_{jm^{\prime}}(x)C  \gamma_{5} U_{ii^{\prime}}(x) \gamma_{5} C D^T_{mj^{\prime}}(x)C \gamma_5  C_{nl^\prime}(x)\gamma_5CU^T_{kk^\prime}(x)C\right]  \nonumber\\
 &&\left.+{\rm Tr}\left[ \gamma_{5} C_{ln^{\prime}}(x)\gamma_{5} CD^T_{jm^{\prime}}(x)C  \gamma_{5} U_{ik^{\prime}}(x) \gamma_{5} C C^T_{nl^{\prime}}(x)C \gamma_5  D_{mj^\prime}(x)\gamma_5CU^T_{ki^\prime}(x)C\right]  \right\}\, ,
\end{eqnarray}
 where the $U_{ij}(x)$, $D_{ij}(x)$ and $C_{ij}(x)$ are the full  $u$, $d$ and $c$ quark propagators, respectively \cite{Reinders85,Pascual-1984}, the $U_{ij}(x)$ and $D_{ij}(x)$ can be written as $S_{ij}(x)$,
\begin{eqnarray}
S_{ij}(x)&=& \frac{i\delta_{ij}\!\not\!{x}}{ 2\pi^2x^4}
 -\frac{\delta_{ij}\langle
\bar{q}q\rangle}{12}  -\frac{\delta_{ij}x^2\langle \bar{q}g_s\sigma Gq\rangle}{192} -\frac{ig_s G^{a}_{\alpha\beta}t^a_{ij}(\!\not\!{x}
\sigma^{\alpha\beta}+\sigma^{\alpha\beta} \!\not\!{x})}{32\pi^2x^2}\nonumber\\
&& -\frac{1}{8}\langle\bar{q}_j\sigma^{\mu\nu}q_i \rangle \sigma_{\mu\nu} +\cdots \, ,
\end{eqnarray}
\begin{eqnarray}
C_{ij}(x)&=&\frac{i}{(2\pi)^4}\int d^4k e^{-ik \cdot x} \left\{
\frac{\delta_{ij}}{\!\not\!{k}-m_c}
-\frac{g_sG^n_{\alpha\beta}t^n_{ij}}{4}\frac{\sigma^{\alpha\beta}(\!\not\!{k}+m_c)+(\!\not\!{k}+m_c)
\sigma^{\alpha\beta}}{(k^2-m_c^2)^2}\right.\nonumber\\
&&\left. -\frac{g_s^2 (t^at^b)_{ij} G^a_{\alpha\beta}G^b_{\mu\nu}(f^{\alpha\beta\mu\nu}+f^{\alpha\mu\beta\nu}+f^{\alpha\mu\nu\beta}) }{4(k^2-m_c^2)^5}+\cdots\right\} \, , \end{eqnarray}
\begin{eqnarray}
f^{\lambda\alpha\beta}&=&(\!\not\!{k}+m_c)\gamma^\lambda(\!\not\!{k}+m_c)\gamma^\alpha(\!\not\!{k}+m_c)\gamma^\beta(\!\not\!{k}+m_c)\, ,\nonumber\\
f^{\alpha\beta\mu\nu}&=&(\!\not\!{k}+m_c)\gamma^\alpha(\!\not\!{k}+m_c)\gamma^\beta(\!\not\!{k}+m_c)\gamma^\mu(\!\not\!{k}+m_c)\gamma^\nu(\!\not\!{k}+m_c)\, ,
\end{eqnarray}
and  $t^n=\frac{\lambda^n}{2}$, the $\lambda^n$ is the Gell-Mann matrix \cite{Reinders85}.
Then we compute  the integrals both in  coordinate space and in momentum space,  and obtain the correlation function $\Pi(p)$ at the quark level, therefore the QCD spectral density
 through dispersion relation.
 In Eq.(7), we retain the term $\langle\bar{q}_j\sigma_{\mu\nu}q_i \rangle$  originates  from the Fierz rearrangement  of the $\langle q_i \bar{q}_j\rangle$ to  absorb the gluons  emitted from other quark lines to form $\langle\bar{q}_j g_s G^a_{\alpha\beta} t^a_{mn}\sigma_{\mu\nu} q_i \rangle$   so as to extract the mixed condensate  $\langle\bar{q}g_s\sigma G q\rangle$ and squared mixed condensate  $\langle\bar{q}g_s\sigma G q\rangle^2$, which play an important role in determining the Borel window, see the typical Feynman diagrams shown in Figs.1-2. It is straightforward but very difficult to calculate those diagrams.

 In this article, we carry out the
operator product expansion to the vacuum condensates  up to dimension-16, and take into account the vacuum condensates which are
vacuum expectations  of the operators  of the orders $\mathcal{O}( \alpha_s^{k})$ with $k\leq 1$ consistently.
The condensates $\langle g_s^3 GGG\rangle$, $\langle \frac{\alpha_s GG}{\pi}\rangle^2$,
 $\langle \frac{\alpha_s GG}{\pi}\rangle\langle \bar{q} g_s \sigma Gq\rangle$ have the dimensions 6, 8, 9, respectively,  but they are   the vacuum expectations
of the operators of the order    $\mathcal{O}( \alpha_s^{3/2})$, $\mathcal{O}(\alpha_s^2)$, $\mathcal{O}( \alpha_s^{3/2})$, respectively, and discarded. Furthermore,
the condensates  $\langle \bar{q}q\rangle\langle \frac{\alpha_s}{\pi}GG\rangle$,
$\langle \bar{q}q\rangle^2\langle \frac{\alpha_s}{\pi}GG\rangle$, $\langle \bar{q}q\rangle^3\langle \frac{\alpha_s}{\pi}GG\rangle$  have dimensions $7$, $10$, $13$, respectively, and they are   the vacuum expectations
of the operators of the order    $\mathcal{O}( \alpha_s)$, however, they play a minor important role,  and neglected \cite{WangHuang-3900,Wang-4660-2014,WangHuang-NPA-2014,WangHuang-mole}.

\begin{figure}
 \centering
 \includegraphics[totalheight=3.5cm,width=10cm]{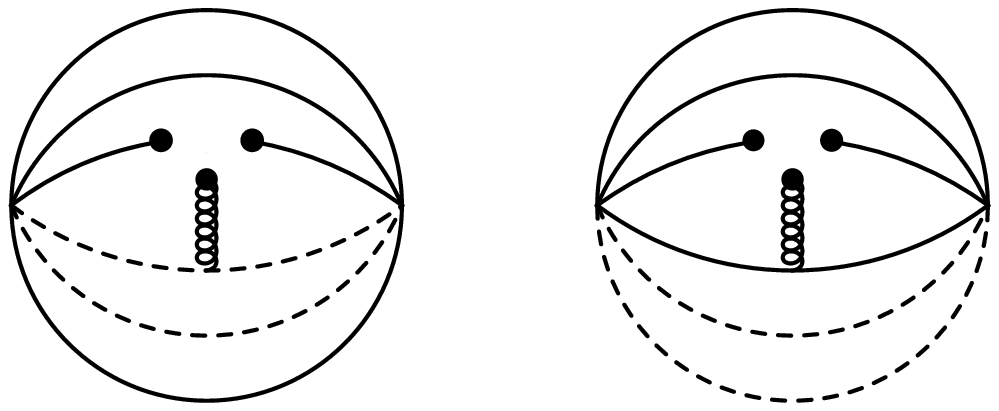}
    \caption{The diagrams contribute  to the mixed condensate $\langle\bar{q}g_s\sigma Gq\rangle$ from the terms $\langle\bar{q}_j\sigma_{\mu\nu}q_i \rangle$. Other
diagrams obtained by interchanging of the heavy quark lines (dashed lines) or light quark lines (solid lines) are implied. }
\end{figure}

\begin{figure}
 \centering
 \includegraphics[totalheight=3.5cm,width=15cm]{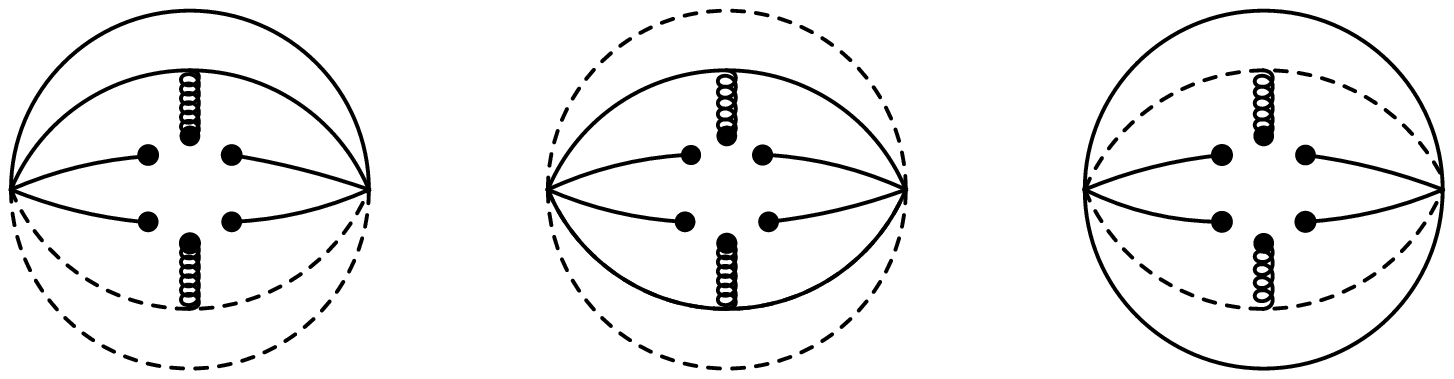}
    \caption{The diagrams contribute  to the squared mixed condensate $\langle\bar{q}g_s\sigma Gq\rangle^2$ from the terms $\langle\bar{q}_j\sigma_{\mu\nu}q_i \rangle$. Other
diagrams obtained by interchanging of the heavy quark lines (dashed lines) or light quark lines (solid lines) are implied. }
\end{figure}

Once the QCD spectral density is obtained, we can  take the
quark-hadron duality and perform Borel transform  with respect to
the variable $P^2=-p^2$ to obtain  the following QCD sum rule,
\begin{eqnarray}
\lambda^2_{Z}\, \exp\left(-\frac{M^2_{Z}}{T^2}\right)= \int_{4m_c^2}^{s_0} ds\, \rho(s) \, \exp\left(-\frac{s}{T^2}\right) \, ,
\end{eqnarray}
where
\begin{eqnarray}
\rho(s)&=&\rho_{0}(s)+\rho_{3}(s)+\rho_{4}(s)+\rho_{5}(s)+\rho_{6}(s)+\rho_{8}(s)+\rho_{9}(s)+\rho_{10}(s)+\rho_{11}(s)+\rho_{12}(s)\nonumber\\
&&+\rho_{13}(s)+\rho_{14}(s)+\rho_{16}(s)\, ,
\end{eqnarray}

\begin{eqnarray}
\rho_0(s)&=&\frac{1}{183500800\pi^{10}}\int_{y_i}^{y_f}dy \int_{z_i}^{1-y}dz\,yz(1-y-z)^5\left(s-\overline{m}_c^2\right)^6\left(9s-2\overline{m}_c^2\right) \nonumber\\
&&+\frac{m_c^2}{235929600\pi^{10}}\int_{y_i}^{y_f}dy \int_{z_i}^{1-y}dz\,(1-y-z)^5\left(s-\overline{m}_c^2\right)^6 \, ,
\end{eqnarray}

\begin{eqnarray}
\rho_3(s)&=&\frac{m_c\langle\bar{q}q\rangle}{491520\pi^{8}}\int_{y_i}^{y_f}dy \int_{z_i}^{1-y}dz\,(y+z)(1-y-z)^4\left(s-\overline{m}_c^2\right)^4\left(7s-2\overline{m}_c^2\right)   \, ,
\end{eqnarray}

\begin{eqnarray}
\rho_4(s)&=&-\frac{m_c^2}{7864320\pi^{8}}\langle \frac{\alpha_sGG}{\pi}\rangle\int_{y_i}^{y_f}dy \int_{z_i}^{1-y}dz\,\left(\frac{z}{y^2}+\frac{y}{z^2} \right)(1-y-z)^5\left(s-\overline{m}_c^2\right)^3\left(3s-\overline{m}_c^2\right)    \nonumber\\
&&-\frac{m_c^4}{35389440\pi^{8}}\langle \frac{\alpha_sGG}{\pi}\rangle\int_{y_i}^{y_f}dy \int_{z_i}^{1-y}dz\,\left(\frac{1}{y^3}+\frac{1}{z^3} \right)(1-y-z)^5\left(s-\overline{m}_c^2\right)^3    \nonumber\\
&&+\frac{m_c^2}{47185920\pi^{8}}\langle \frac{\alpha_sGG}{\pi}\rangle\int_{y_i}^{y_f}dy \int_{z_i}^{1-y}dz\,\left(\frac{1}{y^2}+\frac{1}{z^2} \right)(1-y-z)^5\left(s-\overline{m}_c^2\right)^4    \nonumber\\
&&-\frac{1}{251658240\pi^{8}}\langle \frac{\alpha_sGG}{\pi}\rangle\int_{y_i}^{y_f}dy \int_{z_i}^{1-y}dz\,(1-y-z)^5\left(s-\overline{m}_c^2\right)^4\left(7s-2\overline{m}_c^2\right)    \nonumber\\
&&+\frac{29}{251658240\pi^{8}}\langle \frac{\alpha_sGG}{\pi}\rangle\int_{y_i}^{y_f}dy \int_{z_i}^{1-y}dz\,(y+z)(1-y-z)^4\left(s-\overline{m}_c^2\right)^4\left(7s-2\overline{m}_c^2\right)    \nonumber\\
&&+\frac{19}{62914560\pi^{8}}\langle \frac{\alpha_sGG}{\pi}\rangle\int_{y_i}^{y_f}dy \int_{z_i}^{1-y}dz\,yz(1-y-z)^3\left(s-\overline{m}_c^2\right)^4\left(7s-2\overline{m}_c^2\right)    \nonumber\\
&&-\frac{m_c^2}{188743680\pi^{8}}\langle \frac{\alpha_sGG}{\pi}\rangle\int_{y_i}^{y_f}dy \int_{z_i}^{1-y}dz\, \frac{(1-y-z)^5}{yz}\left(s-\overline{m}_c^2\right)^4     \nonumber\\
&&+\frac{11}{150994944\pi^{8}}\langle \frac{\alpha_sGG}{\pi}\rangle\int_{y_i}^{y_f}dy \int_{z_i}^{1-y}dz\, \overline{m}_c^2 \, (1-y-z)^4\left(s-\overline{m}_c^2\right)^4     \nonumber\\
&&+\frac{17m_c^2}{37748736\pi^{8}}\langle \frac{\alpha_sGG}{\pi}\rangle\int_{y_i}^{y_f}dy \int_{z_i}^{1-y}dz\,  (1-y-z)^3\left(s-\overline{m}_c^2\right)^4     \, ,
\end{eqnarray}

\begin{eqnarray}
\rho_5(s)&=&- \frac{91m_c\langle\bar{q}g_s\sigma Gq\rangle}{4718592\pi^{8}}\int_{y_i}^{y_f}dy \int_{z_i}^{1-y}dz\, (y+z)(1-y-z)^3\left(s-\overline{m}_c^2\right)^3\left(3s-\overline{m}_c^2\right)  \nonumber\\
&&+\frac{11m_c\langle\bar{q}g_s\sigma Gq\rangle}{18874368\pi^{8}}\int_{y_i}^{y_f}dy \int_{z_i}^{1-y}dz\,(1-y-z)^4\left(s-\overline{m}_c^2\right)^3\left(3s-\overline{m}_c^2\right)  \nonumber\\
&&+\frac{13m_c\langle\bar{q}g_s\sigma Gq\rangle}{3145728\pi^{8}}\int_{y_i}^{y_f}dy \int_{z_i}^{1-y}dz\,\left(\frac{z}{y}+\frac{y}{z} \right)(1-y-z)^4\left(s-\overline{m}_c^2\right)^3\left(3s-\overline{m}_c^2\right)  \, ,
\end{eqnarray}

\begin{eqnarray}
\rho_6(s)&=&\frac{7\langle\bar{q}q\rangle^2}{18432\pi^{6}}\int_{y_i}^{y_f}dy \int_{z_i}^{1-y}dz\,yz(1-y-z)^2\left(s-\overline{m}_c^2\right)^3\left(3s-\overline{m}_c^2\right)  \nonumber\\
&&+\frac{7m_c^2\langle\bar{q}q\rangle^2}{9216\pi^{6}}\int_{y_i}^{y_f}dy \int_{z_i}^{1-y}dz\,(1-y-z)^2\left(s-\overline{m}_c^2\right)^3  \, ,
\end{eqnarray}

\begin{eqnarray}
\rho_8(s)&=&-\frac{119\langle\bar{q}q\rangle\langle\bar{q}g_s\sigma Gq\rangle }{147456\pi^{6}}\int_{y_i}^{y_f}dy \int_{z_i}^{1-y}dz\,yz(1-y-z)\left(s-\overline{m}_c^2\right)^2\left(5s-2\overline{m}_c^2\right)  \nonumber\\
&&-\frac{119m_c^2\langle\bar{q}q\rangle\langle\bar{q}g_s\sigma Gq\rangle }{49152\pi^{6}}\int_{y_i}^{y_f}dy \int_{z_i}^{1-y}dz\,(1-y-z)\left(s-\overline{m}_c^2\right)^2  \nonumber\\
&&+\frac{19\langle\bar{q}q\rangle\langle\bar{q}g_s\sigma Gq\rangle }{393216\pi^{6}}\int_{y_i}^{y_f}dy \int_{z_i}^{1-y}dz\,(y+z)(1-y-z)^2\left(s-\overline{m}_c^2\right)^2\left(5s-2\overline{m}_c^2\right)  \nonumber\\
&&+\frac{107\langle\bar{q}q\rangle\langle\bar{q}g_s\sigma Gq\rangle }{196608\pi^{6}}\int_{y_i}^{y_f}dy \int_{z_i}^{1-y}dz\,\overline{m}_c^2\,(1-y-z)^2\left(s-\overline{m}_c^2\right)^2  \, ,
\end{eqnarray}

\begin{eqnarray}
\rho_{9}(s)&=&\frac{m_c\langle\bar{q}q\rangle^3 }{96\pi^{4}}\int_{y_i}^{y_f}dy \int_{z_i}^{1-y}dz\,(y+z)(1-y-z)\left(s-\overline{m}_c^2\right)\left(2s-\overline{m}_c^2\right)  \, ,
\end{eqnarray}

\begin{eqnarray}
\rho_{10}(s)&=& \frac{ 253\langle\bar{q}g_s\sigma Gq\rangle^2 }{393216\pi^{6}}\int_{y_i}^{y_f}dy \int_{z_i}^{1-y}dz\, yz \left(s-\overline{m}_c^2\right)\left(2s-\overline{m}_c^2\right)  \nonumber\\
&&+\frac{ 253m_c^2\langle\bar{q}g_s\sigma Gq\rangle^2 }{393216\pi^{6}}\int_{y_i}^{y_f}dy \int_{z_i}^{1-y}dz  \left(s-\overline{m}_c^2\right)  \nonumber\\
&&-\frac{161 \langle\bar{q}g_s\sigma Gq\rangle^2 }{1048576\pi^{6}}\int_{y_i}^{y_f}dy \int_{z_i}^{1-y}dz\, (y+z)(1-y-z) \left(s-\overline{m}_c^2\right)\left(2s-\overline{m}_c^2\right)  \nonumber\\
&&-\frac{2713 \langle\bar{q}g_s\sigma Gq\rangle^2 }{4718592\pi^{6}}\int_{y_i}^{y_f}dy \int_{z_i}^{1-y}dz\, \overline{m}_c^2\,(1-y-z) \left(s-\overline{m}_c^2\right)  \nonumber\\
&&+\frac{7 m_c^2\langle\bar{q}g_s\sigma Gq\rangle^2 }{32768\pi^{6}}\int_{y_i}^{y_f}dy \int_{z_i}^{1-y}dz \frac{(1-y-z)^2}{yz} \left(s-\overline{m}_c^2\right)  \nonumber\\
&&+\frac{329 \langle\bar{q}g_s\sigma Gq\rangle^2 }{18874368\pi^{6}}\int_{y_i}^{y_f}dy \int_{z_i}^{1-y}dz\, (1-y-z)^2 \left(s-\overline{m}_c^2\right)\left(2s-\overline{m}_c^2\right)  \, ,
\end{eqnarray}

\begin{eqnarray}
\rho_{11}(s)&=&-\frac{ 283m_c \langle\bar{q}g_s\sigma Gq\rangle \langle\bar{q} q\rangle^2 }{36864\pi^{4}}\int_{y_i}^{y_f}dy \int_{z_i}^{1-y}dz\,(y+z) \left(3s-2\overline{m}_c^2\right)  \nonumber\\
&&+\frac{ 29m_c \langle\bar{q}g_s\sigma Gq\rangle \langle\bar{q} q\rangle^2 }{9216\pi^{4}}\int_{y_i}^{y_f}dy \int_{z_i}^{1-y}dz\left(\frac{z}{y}+\frac{y}{z} \right)(1-y-z) \left(3s-2\overline{m}_c^2\right)  \nonumber\\
&&+\frac{ 19m_c \langle\bar{q}g_s\sigma Gq\rangle \langle\bar{q} q\rangle^2 }{12288\pi^{4}}\int_{y_i}^{y_f}dy \int_{z_i}^{1-y}dz\,(1-y-z) \left(3s-2\overline{m}_c^2\right)  \, ,
\end{eqnarray}

\begin{eqnarray}
\rho_{12}(s)&=&\frac{  \langle\bar{q} q\rangle^4 }{864\pi^{2}}\int_{y_i}^{y_f}dy  \,y(1-y) \left(3s-2\widetilde{m}_c^2\right)+\frac{ m_c^2 \langle\bar{q} q\rangle^4 }{96\pi^{2}}\int_{y_i}^{y_f}dy\, ,
\end{eqnarray}

\begin{eqnarray}
\rho_{13}(s)&=&-\frac{ 355m_c \langle\bar{q}q\rangle \langle\bar{q}g_s\sigma G q\rangle^2 }{110592\pi^{4}}\int_{y_i}^{y_f}dy \int_{z_i}^{1-y}dz\left( \frac{z}{y}+\frac{y}{z}\right) \left[1+\frac{s}{2} \,\delta\left(s-\overline{m}_c^2\right) \right] \nonumber\\
&&-\frac{ 313m_c \langle\bar{q}q\rangle \langle\bar{q}g_s\sigma G q\rangle^2 }{196608\pi^{4}}\int_{y_i}^{y_f}dy \int_{z_i}^{1-y}dz \left[1+\frac{s}{2} \,\delta\left(s-\overline{m}_c^2\right) \right] \nonumber\\
&&+\frac{ 467m_c \langle\bar{q}q\rangle \langle\bar{q}g_s\sigma G q\rangle^2 }{884736\pi^{4}}\int_{y_i}^{y_f}dy \int_{z_i}^{1-y}dz\left( \frac{1}{y}+\frac{1}{z}\right)(1-y-z) \left[1+\frac{s}{2} \,\delta\left(s-\overline{m}_c^2\right) \right] \nonumber\\
&&+\frac{139m_c \langle\bar{q}q\rangle \langle\bar{q}g_s\sigma G q\rangle^2 }{36864\pi^{4}}\int_{y_i}^{y_f}dy  \left[1+\frac{s}{2} \,\delta\left(s-\widetilde{m}_c^2\right) \right] \, ,
\end{eqnarray}

\begin{eqnarray}
\rho_{14}(s)&=&-\frac{  \langle\bar{q}q\rangle^3 \langle\bar{q}g_s\sigma G q\rangle }{432\pi^{2}}\int_{y_i}^{y_f}dy \,y(1-y) \left[3+\left(\frac{13}{2}+\frac{5s}{T^2}\right)\,s\, \,\delta\left(s-\widetilde{m}_c^2\right)  \right] \nonumber\\
&&+\frac{11  \langle\bar{q}q\rangle^3 \langle\bar{q}g_s\sigma G q\rangle }{27648\pi^{2}}\int_{y_i}^{y_f}dy     +\frac{185   \langle\bar{q}q\rangle^3 \langle\bar{q}g_s\sigma G q\rangle }{55296\pi^{2}}\int_{y_i}^{y_f}dy     \,s\,\delta\left(s-\widetilde{m}_c^2\right) \, ,
\end{eqnarray}

\begin{eqnarray}
\rho_{16}(s)&=&\frac{  \langle\bar{q}q\rangle^2 \langle\bar{q}g_s\sigma G q\rangle^2 }{384\pi^{2}}\int_{y_i}^{y_f}dy  \,y(1-y)\left( 1+\frac{s}{T^2}+\frac{s^2}{2T^4}+\frac{5s^3}{3T^6}\right)\,\delta\left(s-\widetilde{m}_c^2\right) \nonumber\\
&&-\frac{1255  \langle\bar{q}q\rangle^2 \langle\bar{q}g_s\sigma G q\rangle^2 }{5308416\pi^{2}}\int_{y_i}^{y_f}dy   \,\delta\left(s-\widetilde{m}_c^2\right) \nonumber\\
&&+\frac{151  \langle\bar{q}q\rangle^2 \langle\bar{q}g_s\sigma G q\rangle^2 }{196608\pi^{2}}\int_{y_i}^{y_f}dy   \frac{s}{T^2} \,\delta\left(s-\widetilde{m}_c^2\right) \nonumber\\
&&-\frac{185  \langle\bar{q}q\rangle^2 \langle\bar{q}g_s\sigma G q\rangle^2 }{73728\pi^{2}}\int_{y_i}^{y_f}dy \,  \frac{s^2}{T^4}\, \delta\left(s-\widetilde{m}_c^2\right) \, ,
\end{eqnarray}
  $y_{f}=\frac{1+\sqrt{1-4m_c^2/s}}{2}$,
$y_{i}=\frac{1-\sqrt{1-4m_c^2/s}}{2}$, $z_{i}=\frac{y
m_c^2}{y s -m_c^2}$, $\overline{m}_c^2=\frac{(y+z)m_c^2}{yz}$,
$ \widetilde{m}_c^2=\frac{m_c^2}{y(1-y)}$, $\int_{y_i}^{y_f}dy \to \int_{0}^{1}dy$, $\int_{z_i}^{1-y}dz \to \int_{0}^{1-y}dz$ when the $\delta$ functions $\delta\left(s-\overline{m}_c^2\right)$ and $\delta\left(s-\widetilde{m}_c^2\right)$ appear,
   the $s_0$ is the continuum threshold parameter.

 We derive  Eq.(10) with respect to  $\tau=\frac{1}{T^2}$, then eliminate the
 pole residue  $\lambda_{Z}$ to obtain the QCD sum rule for the mass,
 \begin{eqnarray}
 M^2_{Z}= \frac{-\frac{d}{d \tau } \int_{4m_c^2}^{s_0} ds\,\rho(s)\,e^{-\tau s}}{\int_{4m_c^2}^{s_0} ds \,\rho(s)\,e^{-\tau s}}\, .
\end{eqnarray}

\section{Numerical results and discussions}

We take  the standard values of the vacuum condensates $\langle
\bar{q}q \rangle=-(0.24\pm 0.01\, \rm{GeV})^3$,   $\langle
\bar{q}g_s\sigma G q \rangle=m_0^2\langle \bar{q}q \rangle$,
$m_0^2=(0.8 \pm 0.1)\,\rm{GeV}^2$, $\langle \frac{\alpha_s
GG}{\pi}\rangle=(0.33\,\rm{GeV})^4 $    at the energy scale  $\mu=1\, \rm{GeV}$
\cite{SVZ79,Reinders85,Colangelo-Review}, and choose the $\overline{MS}$ mass $m_{c}(m_c)=(1.275\pm0.025)\,\rm{GeV}$  from the Particle Data Group \cite{PDG}.
Moreover, we take into account the energy-scale dependence of  the input parameters,
\begin{eqnarray}
\langle\bar{q}q \rangle(\mu)&=&\langle\bar{q}q \rangle(Q)\left[\frac{\alpha_{s}(Q)}{\alpha_{s}(\mu)}\right]^{\frac{4}{9}}\, ,\nonumber\\
 \langle\bar{q}g_s \sigma Gq \rangle(\mu)&=&\langle\bar{q}g_s \sigma Gq \rangle(Q)\left[\frac{\alpha_{s}(Q)}{\alpha_{s}(\mu)}\right]^{\frac{2}{27}}\, ,\nonumber\\
m_c(\mu)&=&m_c(m_c)\left[\frac{\alpha_{s}(\mu)}{\alpha_{s}(m_c)}\right]^{\frac{12}{25}} \, ,\nonumber\\
\alpha_s(\mu)&=&\frac{1}{b_0t}\left[1-\frac{b_1}{b_0^2}\frac{\log t}{t} +\frac{b_1^2(\log^2{t}-\log{t}-1)+b_0b_2}{b_0^4t^2}\right]\, ,
\end{eqnarray}
  where $t=\log \frac{\mu^2}{\Lambda^2}$, $b_0=\frac{33-2n_f}{12\pi}$, $b_1=\frac{153-19n_f}{24\pi^2}$, $b_2=\frac{2857-\frac{5033}{9}n_f+\frac{325}{27}n_f^2}{128\pi^3}$,  $\Lambda=213\,\rm{MeV}$, $296\,\rm{MeV}$  and  $339\,\rm{MeV}$ for the flavors  $n_f=5$, $4$ and $3$, respectively  \cite{PDG}, and evolve all the input parameters to the optimal energy scale  $\mu$ to extract the mass of the $Z^{++}_{cc}$.

In Refs.\cite{WangHuang-3900,Wang-4660-2014,WangHuang-NPA-2014,WangHuang-mole}, we study the acceptable energy scales of the QCD spectral densities  for the hidden-charm (hidden-bottom) tetraquark states and molecular states   in the QCD sum rules  for the first time,  and suggest an  empirical formula  $\mu=\sqrt{M^2_{X/Y/Z}-(2{\mathbb{M}}_Q)^2}$ to determine  the optimal  energy scales. The energy scale formula enhances the pole contributions remarkably and works well.
The  energy scale formula also works well in studying the hidden-charm pentaquark states \cite{WangPc}. In this article, we study the diquark-diquark-diquark type hexaquark state, the basic constituent are also diquarks, just like in the case of the diquark-antidiquark type tetraquark  states \cite{WangHuang-3900,Wang-4660-2014,WangHuang-NPA-2014}. So we extend our previous work to study the hexaquark state by taking  the energy scale formula $\mu=\sqrt{M^2_{X/Y/Z}-(2{\mathbb{M}}_c)^2}$ with the updated value ${\mathbb{M}}_c=1.82\,\rm{GeV}$ as a constraint to obey \cite{Wang-1601}.

Experimentally, there is no candidate for the doubly charged hexaquark state $Z_{cc}^{++}$ with the symbolic quark structure $uuddcc$.
In the  scenario of tetraquark  states, the QCD sum rules indicate that the $Z_c(3900)$ and $Z(4430)$ can be tentatively assigned to be the ground state and the first radial excited state of the axialvector tetraquark states, respectively \cite{Wang4430}, the $Y(3915)$ and $X(4500)$ can be tentatively assigned to be the ground state and the first radial excited state of the scalar tetraquark states, respectively \cite{Wang-3915-CgmCgm}. The energy gap between the ground state and the first radial excited state of the hidden-charm tetraquark states is about $0.6\,\rm{GeV}$. Now we suppose the energy gap between the ground state and the first radial excited state of the doubly charmed  hexaquark states is about $0.6\,\rm{GeV}$, and   tentatively take the continuum threshold parameter  to be   $\sqrt{s_0}=M_{Z}+(0.4\sim0.6)\,\rm{GeV}$, which also serves as a constraint to obey.

We search for the optimal Borel parameter and continuum threshold parameter to satisfy the two criteria (pole dominance and convergence of the operator product
expansion) of the QCD sum rules, and obtain the values $T^2=(5.3-5.7)\,\rm{GeV}^2$ and $\sqrt{s_0}=(7.1\pm0.1)\,\rm{GeV}$ for the energy scale $\mu=5.5\,\rm{GeV}$,  the predicted mass satisfies the energy scale formula and the continuum threshold parameter satisfies our naive expectation.
The pole contribution is about $(26-41)\%$, the pole dominance condition is not satisfied, see Fig.3. In fact, if we do not use the energy scale formula, the pole contribution is much smaller.  In Fig.4, we plot the contributions of the vacuum condensates in the operator product expansion with variations of the Borel parameter $T^2$ for the value $\sqrt{s_0}=7.1\,\rm{GeV}$. From the figure, we can see that the vacuum condensates of dimensions $10$, $12$, $13$, $14$, $16$ play a minor important role in the Borel window, the operator product expansion is well convergent.  In calculations, we observe that the integral $\int_{4m_c^2}^{s_0} ds\, \rho(s) \, \exp\left(-\frac{s}{T^2}\right)$ is negative at the region $T^2<4\,\rm{GeV}^2$ for $\sqrt{s_0}=7.1\,\rm{GeV}$. Although the  vacuum condensates of dimensions $10$, $12$, $13$, $14$, $16$ play a minor important role in the Borel window, they play an important role in determining the Borel window. In Fig.5, we plot the mass with variation of the Borel parameter $T^2$ by taking into account the vacuum condensates up to dimensions 16 and 10, respectively. From the figure, we can see that the predicted mass decreases  monotonously with increase of the Borel parameter $T^2$ for the truncation $n\leq 10$, there appears no platform.

We take  into account all uncertainties of the input parameters,
and obtain the values of the mass and pole residue of
 the    $Z_{cc}^{++}$, which are  shown explicitly in Figs.6-7,
\begin{eqnarray}
M_{Z}&=&6.60^{+0.12}_{-0.09}\,\rm{GeV} \, ,  \nonumber\\
\lambda_{Z}&=&7.64^{+1.17}_{-1.05}\times 10^{-3}\,\rm{GeV}^8 \,   .
\end{eqnarray}
From Figs.6-7, we can see that there appear platforms at the Borel window $T^2=(5.3-5.7)\,\rm{GeV}^2$,  no platform can be obtained at the value $T^2<5.2\,\rm{GeV}^2$.
The predicted mass $M_{Z}=6.60^{+0.12}_{-0.09}\,\rm{GeV}$ lies above the thresholds $\Sigma_c(2455)\Sigma_c(2455)$ and $\Sigma_c(2520)\Sigma_c(2520)$, the decays to the  charmed-baryon pairs $\Sigma_c(2455)\Sigma_c(2455)$ and $\Sigma_c(2520)\Sigma_c(2520)$ are  Okubo-Zweig-Iizuka  super-allowed, we can search for the $Z_{cc}^{++}$ in those decay channels. The diquark-diquark-diquark type hexaquark state is not a baryon-baryon type dibaryon \cite{MOka-94} or a baryon-antibaryon type baryonium \cite{Wang-1835}, whose masses lie near the corresponding thresholds. In the QCD sum rules for the dibaryon or baryonium, the pole dominance is also failed to satisfy. In Ref.\cite{Di-baryon-QM}, it is observed that no stable hexaquark states exist below the corresponding two-baryon thresholds based on a simple potential quark model. In the present work, we observe that the scalar hexaquark state lies  far  above the $\Sigma_c(2455)\Sigma_c(2455)$ and $\Sigma_c(2520)\Sigma_c(2520)$ thresholds.

\begin{figure}
 \centering
 \includegraphics[totalheight=6cm,width=8cm]{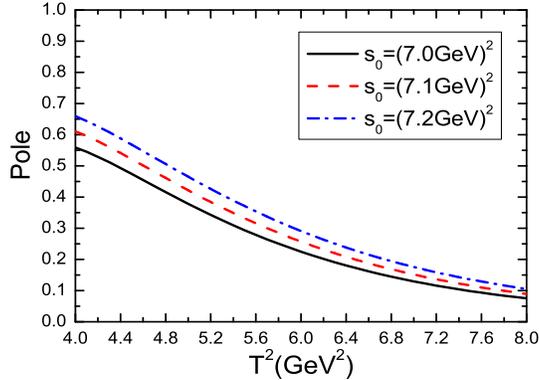}
        \caption{ The pole contribution  of the $Z_{cc}^{++}$  with variation of the Borel parameter $T^2$.  }
\end{figure}

 \begin{figure}
 \centering
 \includegraphics[totalheight=5cm,width=7cm]{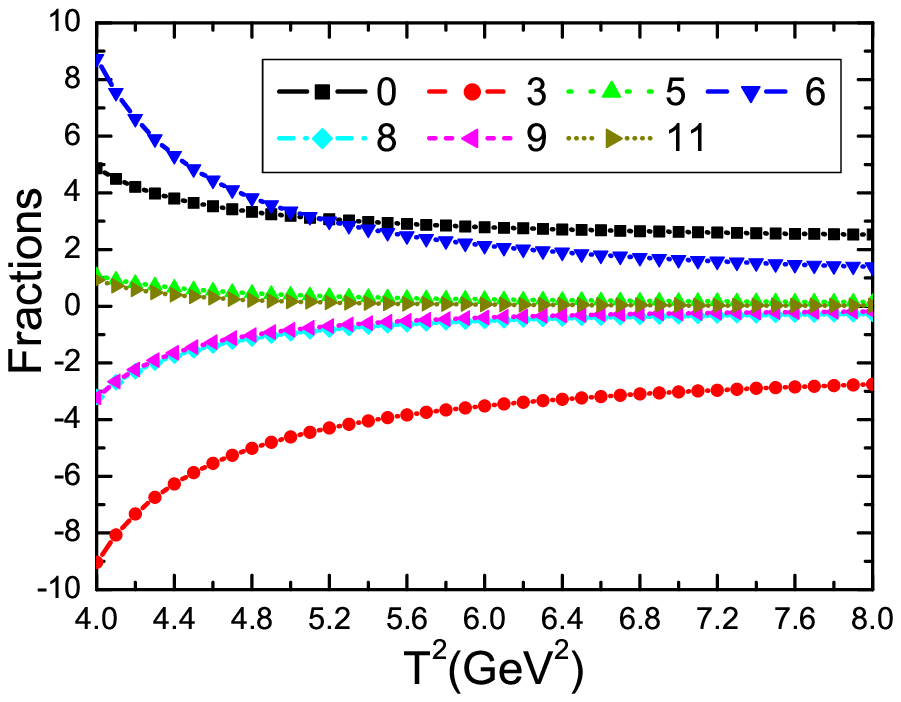}
 \includegraphics[totalheight=5cm,width=7cm]{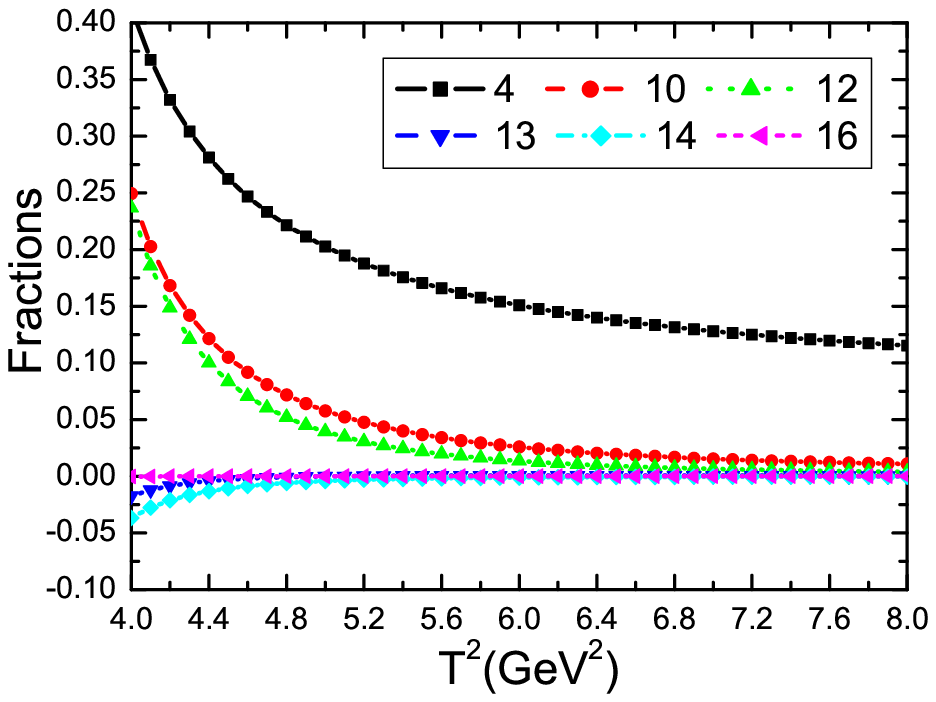}
        \caption{ The contributions of  different terms in the operator product expansion    with variations  of the Borel parameter $T^2$, where the $0$, $3$, $4$, $5$, $6$, $\cdots$ denote the dimensions of the vacuum condensates.  }
\end{figure}

\begin{figure}
 \centering
 \includegraphics[totalheight=6cm,width=8cm]{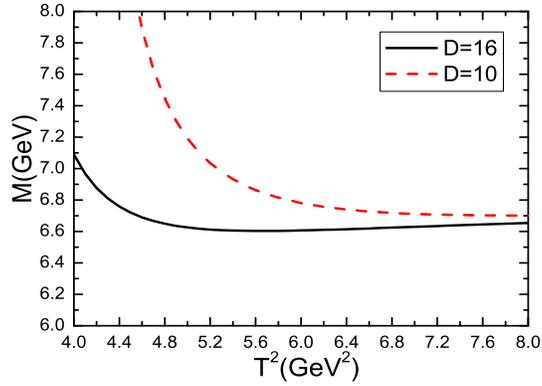}
        \caption{ The mass  of the $Z_{cc}^{++}$  with variation of the Borel parameter $T^2$, where the $D=16$ and $D=10$ denote the truncations in the operator product expansion.  }
\end{figure}

\begin{figure}
 \centering
 \includegraphics[totalheight=6cm,width=8cm]{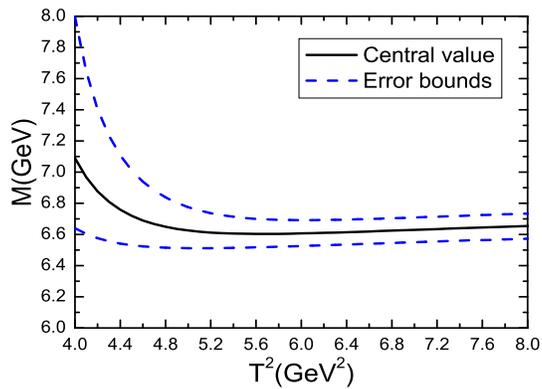}
        \caption{ The mass  of the $Z_{cc}^{++}$  with variation of the Borel parameter $T^2$.  }
\end{figure}

\begin{figure}
 \centering
 \includegraphics[totalheight=6cm,width=8cm]{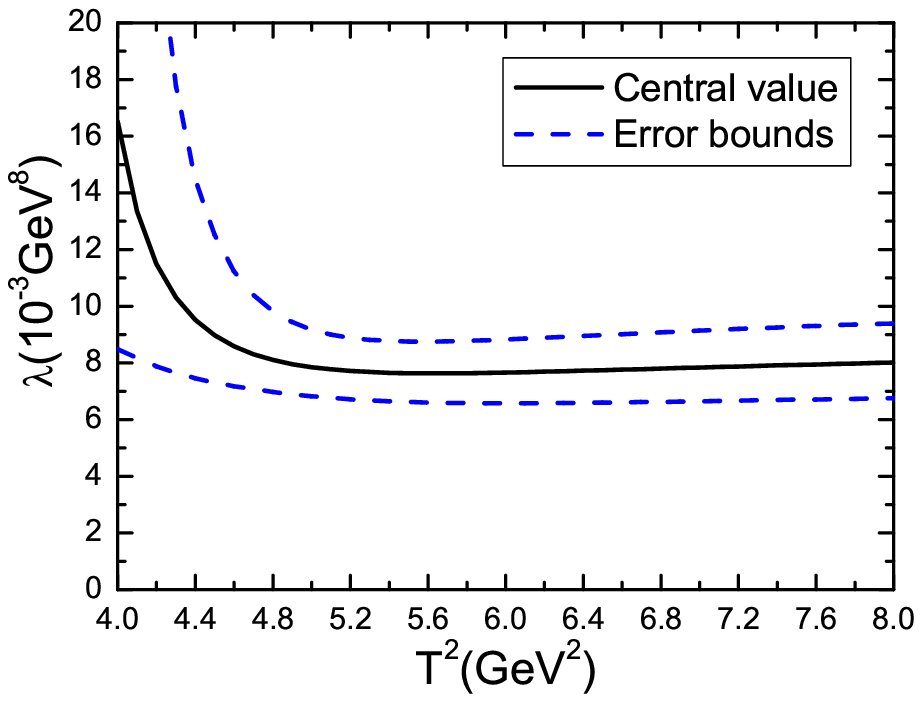}
        \caption{ The  pole residue  of the $Z_{cc}^{++}$ with variation of the Borel parameter $T^2$.  }
\end{figure}

\section{Conclusion}
In this article, we construct the scalar-diquark-scalar-diquark-scalar-diquark type current to interpolate the scalar hexaquark state, and study it with QCD sum rules by carrying out the operator product expansion up to the vacuum condensates of dimension 16. In calculation, we take the energy scale formula as a constraint to determine the energy scale of the QCD spectral density to extract the mass and pole residue. In the Borel window, the operator product expansion is well convergent, while the pole contribution is about $(26-41)\%$.  We obtain the lowest hexaquark mass $M_{Z}=6.60^{+0.12}_{-0.09}\,\rm{GeV}$, which can be confronted to the experimental data in the future, while the predicted pole residue can be used to study the strong decays of the hexaquark state with the three-point QCD sum rules.

\section*{Acknowledgements}
This  work is supported by National Natural Science Foundation, Grant Number 11375063.

\end{document}